# NANDA Adaptive Resolver: Architecture for Dynamic Resolution of AI Agent Names

| John Zinky | Hema Seshadri | Mahesh Lambe | Pradyumna Chari | Ramesh Raskar |
|---|---|---|---|---|
| Akamai Technologies | Akamai Technologies | Unify Dynamics | MIT Media Lab | MIT Media Lab |

## Abstract

AdaptiveResolver is a dynamic microservice architecture designed to address the limitations of static endpoint resolution for AI agent communication in distributed, heterogeneous environments. Unlike traditional DNS or static URLs, AdaptiveResolver enables context-aware, real-time selection of communication endpoints based on factors such as geographic location, system load, agent capabilities, and security threats. Agents advertise their Agent Name and context requirements through Agent Fact cards in an Agent Registry/Index. A requesting Agent discovers a Target Agent using the registry. The Requester Agent can then resolve the Target Agent Name to obtain a tailored communication channel to the agent based on actual environmental context between the agents. The architecture supports negotiation of trust, quality of service, and resource constraints, facilitating flexible, secure, and scalable agent-to-agent interactions that go beyond the classic client-server model. AdaptiveResolver provides a foundation for robust, future-proof agent communication that can evolve with increasing ecosystem complexity.

## Introduction

Analogous to Internet host name resolution using DNS, the AdaptiveResolver service dynamically converts an "Agent Name" to a communication endpoint to that AI agent. For example, the AdaptiveResolver could return an URL, whereas DNS returns an IP address. A major point is that Agent Name resolution happens within a context. A context that could involve properties of both agents and the communication environment between them. The resolution process produces a communication channel "tailored" to that context. So, different clients may get different URLs to the same Agent, if their context is different. A Tailored Response can be reused or cached, only if all entities involved share an overlapping or identical context.

An AI agent advertises its requirements to establish a communication end-point by adding properties to an AgentFacts card. For example, the Agent Name and service geolocation. The card is then published via a NANDA Index ( i.e. a lightweight index/registry for agent discovery and identity), where potential peer agents can discover the agent based on the capabilities published in the card. This paper assumes that the Requester Agent has gone through the Discovery process and has access to the Target's AgentsFacts card.

The simplest form of a communication endpoint is a static URL, which could just be a field in the Agent Facts card. This short cut would avoid the dynamic resolution process described here. However, while static endpoints are adequate for small interactions, past experience with the Internet shows that static endpoints do not scale and are susceptible to DDoS attacks, overload, and flash mobs.

We can learn from the Internet at scale that the simple model that a "DNS host name is resolved into a single IP address" is not true for even moderate size web-applications. For popular hosts, the DNS Authoritative Name Server ([rfc7871](rfc7871)) returns a "Tailored Response" based on the perceived topological or geographic location of the requester, i.e. the requester's context. This allows the communication path between the requester and the target service to be groomed to improve performance, security, and resource consumption. For example, clients may be given an IP address that is physically close to them, cache the content of the latest game download, or spread out the load of a DDoS attack. Logically, this support for communication is as if the service's deployment has been spread out over the network and is not located in one datacenter. DNS has evolved over the decades to support the Internet at scale. We need to use the lessons learned to support AI Agent to AI Agent communication at scale.

Agent to Agent communication has a richer set of deployment opportunities and restrictions than classic web-based interactions of Client to Server systems. First off, both agents are active and interact over a rich physical environment. Both agents may want to participate in negotiations on how they will be used and who to trust. The interaction between agents is more peer-to-peer and session oriented than the typical stateless web transactions and may benefit from services offered by the communication environment. Also, the physical hardware puts limits on how agent functionality can be deployed, but may allow for opportunities to deploy parts of the agents in close proximity. Finally, the Quality of Service and cost requirements have a huge range of possibilities, so that one size does not fit all.

The basic theme of the dynamic resolution architecture is to offer commitment times where the Agents incrementally agree on the conditions under which they will interact, ultimately resulting in a communication channel between a group of agents. The agents need to negotiate requirements for trust and to set expectations. They must establish a specification for expected usage patterns, resource availability, QoS requirements, and cost constraints. From these a communication plan must be optimized and deployed in the physical environment. Any of these steps can be skipped, if all parties agree on them ahead of time.

The architectural goal of dynamic resolution is to allow hooks for all the steps needed to deploy agreeable communication between agents over a physical environment.  But the actual optimization algorithms will not be specified, only motivating examples will be given. These deployment modes will be the subject of future designs and are expected to evolve dramatically over time.  The result is easy implementation for early adopters with a smooth path to evolve into a robust, secure, and scalable future.

This paper starts with an architecture for a simple dynamic resolver system based on DNS. Instead of using raw packets, the system uses REST APIs, and hence can reuse all the security and hosting technologies developed for web-based applications, similar to DoH [[RFC 8484](#)]. Then we will describe the opportunities and restrictions for AI Agent to AI Agent communications, concentrating on the context for which the communication will be deployed. How the communication is set up depends on the agent's desired requirements and the physical restriction of the communication environment. We give some examples of different deployment strategies that should be supported. Finally, we extend the resolver system to add hooks for adaptive resolution, with negotiation of comms requirements, optimization of comms placement, and the actual setting up of the comms channel.

## Dynamic Agent Name Resolver Architecture

Dynamic Resolution is concerned with *how* communication channels are set up between agents and not about discovering agents themselves. The Dynamic Name Resolution mediates interactions between the Requester and Target Agents and the communication environment between them, resulting in scalable communication channels that are secure, efficient and reliable.

Dynamic resolution is concerned with two pieces of information:
**Agent Names** are human readable strings that uniquely identify an agent. A requester can resolve an Agent Name to get a communication end-point to the named agent. Dynamic resolvers may return different end-points to different requesters based on the *context* of the agents and the communication environment between them.

**Agent Facts card,** on the other hand, is a bundle of metadata that is used to discover agents which meet a criteria. Hence, the metadata in an Agent Facts are focused on *what* is the agent's functionality, as opposed to *how* the agent is deployed. Agent Facts are public, where the details of deployment can be kept private until the actual communication channel is set up. Some of the metadata may concern agent name resolution, for example, the Agent's Name and *requirements* and *restrictions* on any requester's context.

Target Agents give these pieces of information to two different systems. Agent Fact cards are registered in the Agent Discovery system, while the Target Agent's name is recorded in an agent Name Space. The Requester gets an appropriate Target Agent Fact card from the Agent Discovery system. The Requester extracts the Agent Name and appends its context. The Requester then uses a Dynamic Resolver system to query the Name Space for the Target Agent Name and set up a communication channel to the Target Agent.

This section explains the Dynamic Resolution Architecture. First we will discuss the syntax of an Agent Path Name. Next we show how the Target Agent records itself into the Name Space. Also, we briefly discuss what the metadata the Target Agent should include in the Agent Facts card to facilitate resolution. Finally, we go through the recursive process the Requester Agent uses to resolve an Agent Name.

## Agent Path Name

Agent Names must be understandable by humans, but rigorous enough to be resolved uniquely by machines. We propose that Agents Names use the lessons learned from DNS names and use hierarchical path names. The Agent Name Space is a hierarchy, The root node is the ID for the name space itself, with subtrees that delegate zones of responsibility. An Agent Name is a leaf in a path name hierarchy.

Notice that Agent Name is independent of the context in which it is resolved. The path name should not muddle identity with communication issues, such as protocol, version, security, and other context metadata. These context issues will be addressed at resolution time, not naming time.

The IETF standard for Uniform Resource Names (URN) [RFC 8141] partitions a name string into the following parts:
1) **Name Prefix** is syntactic sugar that indicates that the string should be interpreted as a Name.
2) **Name Space Identifier** is a unique string for a name space.
3) **Path Name String** to the agent name relative the Name Space

URN name strings have the form:
`urn:<namespace_id>:<path_name_string>`

Name Space Identifier (NID), to be an official URN, must be universally registered in the Internet Assigned Numbers Authority (IANA).   We proposed instead that the NID string be the DNS name for the root name server that is managing the Name Space.  Using a DNS name would reuse the existing DNS name registry, certificate infrastructure, and DNS scaling. The URL to the root name server could be derived by adding https as the protocol and an OpenAPI spec for the different query functions.

The Path Name String is a path down the name hierarchy to a leaf. Each level sub-name is separated by a character such as ":". We prefer not to use "/" as the separator as that will confuse the agent path name with a file path name in a directory. [or maybe file names is a good analogy?]

Let's call our Agent Name a "Uniform Agent Locator"  or UAL. An example name may have the form:
`ual:nanda.mit.edu:lab15:robot42`
`ual:nanda.mit.edu/lab15/robot42`
`agent:nanda.mit.edu:lab15:robot42`
`agent:nanda.mit.edu/lab15/robot42`
`@nanda.mit.edu:lab15:robot42`

Other standard URI styles for an Agent Name have a more verbose form. But while these are familiar and standard, they may have too much syntax for non-programers:

```
# URN namespace identifier ("agent") needs to be registered with IANA
URN -> urn:agent:nanda.mit.edu:lab15:robot42

# URI defines a new scheme "agent" and rest of the syntax is standard
URI -> agent://nanda.mit.edu/lab15/robot42

# URL to a DNS subdomain for a Name Space Server.
# This allows DNS load balancing, similar to "www." for webservers,
URL -> https://agent.nanda.mit.edu?path='lab15:robot42'

# URL to a Fixed DNS Name Space Server
# This allows "/agent" reverse proxy dispatch (nginx)
URL -> https://nanda.mit.edu/agent/lab15/robot42
URL -> https://nanda.mit.edu/agent?path='lab15:robot42'
```

## Recording an Agent Name into a Name Space

The Name Space solves the problem of creating universally unique names for potentially a large number of agents. The Name Space partitions the responsibility and ownership to Intermediate Name Servers. The intermediaries may be owned by different organizations, such as owners of the agent's function, the owner of the deployment infrastructure, or a third party broker. Trust for zones of responsibility are enforced off line when entities register. All registration is done local to the Name Space and not some external organization or government. External trust is relative to the reputation of the Name Space. (See step 1 of figure)

An Agent records its name with the Intermediate Name Server that is responsible for its branch of the path name. The Agent Deployment Record can have metadata about how its instance is deployed and what physical resources it owns and where the resources are located both geographically and topologically. The more flexible the Agent deployment the more tailored the communication can be matched to the context of potential future Requesting Agents and the communication resources between them. (See step 2 of figure)

The last Name Server is responsible for the Agent itself and called the "Authoritative Name Server". It has the task of actually finding or making the communication channel from the Requester to the Target and delivering the end-point (URL) to both Agents.
An Authoritative Name Server is potentially a complicated task that might be outsourced to a third party that supplies communication resources.[Akamai Global Traffic Management] (See step 3 of figure)

At resolution time, the Authoritative Name Server may have many options for how to construct the communication channel and has an opportunity to optimize the channel based on the context.

This optimization task is in the critical path of the resolution process. The Authoritative Name Server must be ready to make a quick but accurate decision. This could involve monitoring the status of target and comms resources and pre-calculating the cost of different scenarios. (See step 4 of figure)

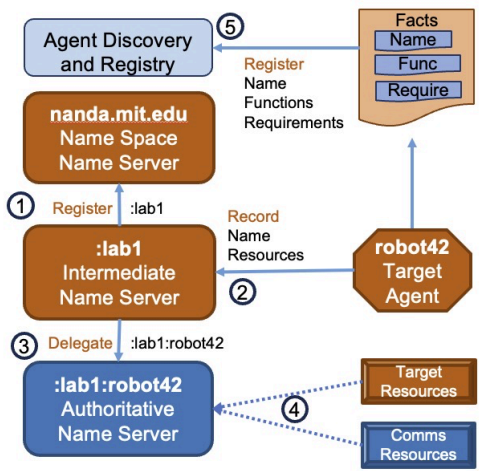

**ual:nanda.mit.edu:lab15:robot42**

1. Intermediate Name Server registers as owner of zone :lab1

2. Agent records its name and resources with Name Server

3. Name Server delegate Agent name to an Authoritative Name Server

4. Authoritative Name Server monitors status of resources

5. Agent register its name, functionality, and comm requirement into Discovery Service

## Registering AgentFacts card in an Agent Registry

Once a Target Agent has created its Agent Name Deployment Record, the agent can publish its capabilities as an Agent Facts card. Most of the metadata in an AgentFacts card will concern the functionality of the agent. But some of the metadata will concern the options for how it can be deployed. The Target Agent can set requirements and restrictions on the context for Requesters. For example, it may require the city where the Requester is located or a CIDR for its topological address [RFC 7871]. The more detailed the demands on context information, the faster the negotiation for trust. If the Requester gives all the context information requested by the Target and does not ask for any new context information from the Target, then the negotiation step can be skipped. The context requirements in the AgentFacts card is a way of passing static metadata from the Target to the Requester. (See step 5 of figure)

## Resolving an Agent Name through a Recursive Resolver

The Agent Name Resolving process starts when the Requester Agent finds a Target Agent for which it wants to establish communication. The Requester Agent has the Target Agent's Agent Facts card and extracts the Agent Name and the Context Requirements metadata. The Requester creates a Resolver Query that includes the Target Agent Name and its own context metadata. (See step 0 of figure)

The Requester has access to a UAL Recursive Resolver Service, which will take care of the process of calling the Name Space to resolve the Agent Name. Like an Internet DNS resolver, this may take many iterations until it has access to the Authoritative Name Server for the name. The Recursive Resolver is doing this work for the Requester and can be implemented as a library in the Requester code or as a standalone server. The Resolver Query is sent from the Requester Agent to the Recursive Resolver. (See step 1 of figure)

The Recursive Resolver uses the Agent Name to extract Name Space ID. It then derives the Name Space Name Server's URL from the ID. Resolver Query is made to the Name Space Name Server with the hope that it is the Authoritative Name Server or knows the URL for the Authoritative Name Server. (See step 2 of figure)

The Name Space Name Server might return a Referral to an Intermediate Name Server that is responsible for the level 1 zone. Since this information is relatively static and has a time-to-live, the referral can be cached by the Recursive Resolver for future requests. (See step 3 of figure)

The Intermediate Name Server sends a referral to the Authoritative Name Server. (See step 4-5) of figure)

The Authoritative Name Server has the Requester context from the Resolver Query. The Target context and comms meta data is in the Agent Deployment Record. (see step 6 of figure)

The Authoritative Name Server quickly checks the status of the comms and target resources and any pre-calculated cost of potential options. It has enough information to optimize the communication channel. It finds or makes the communication channel (see step 7 of figure)

The Authoritative Name Server returns the URL for the Requester end-point (see step 8 of figure)

The Recursive Resolver returns the URL for the Requester end-point (see step 9 of figure)

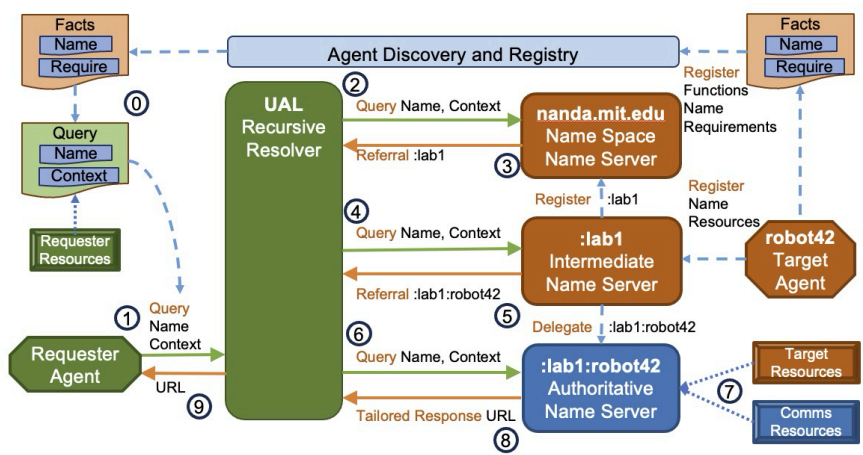

0. Discover Target Agent Name
ual:nanda.mit.edu:lab15:robot42

1. Query Name with Context
2. Query Name Space
3. Refer to Intermediate
4. Query Intermediate
5. Refer to Authoritative
6. Query Authoritative
7. Tailor Response
8. Return Tailored URL
9. Return URL

# Adapting to Environment Context

How inter-agent communication should be deployed depends on the context in which the interaction takes place. If there are multiple options for how to deliver the same functionality, then *adaptive* deployment must choose the appropriate option given the current status of the environment. Different allocations of functions to resources come with different tradeoffs. The agents must agree on which tradeoffs are important and which can be ignored. With more flexibility in the deployment options and a richer understanding of the context, then the deployment will be more scalable, secure, and inexpensive.

In DNS in a client-server scenario, the Authoritative Name Server is responsible for deploying the target server and comms functionality and assuming a simple/dumb communication environment between the returned end-point and the requester. But AI Agents may demand equal participation in the setting up from all the owners of agents and the owners of the comms environment.

The inter-agent context has rich interactions between Agents and the diverse communication environments. To take advantage of these opportunities, the resolution process needs to add more opportunities for interaction between entities. Understanding the requirements and restrictions on the environment is a key piece of knowledge needed for adaptive resolution.

Adaptive deployment needs knowledge about several categories of metadata about the environment context. These are described below:

## AI Agent Implementation

Not all AI agents will be implemented as a monolithic server hosted in a data center with a single origin end-point. Scalable AI agents will be modular, made up of components that work together to implement the agent functionality. Componentization gives flexibility for how these components are deployed over physical resources. Think of AI Agents as amorphous amoebae reaching out with their arms to contact other amoebae. Each component has requirements for different physical resources, such as LLMs need powerful GPUs, vector stores need large amounts of storage, and training data streaming needs huge amounts of bandwidth.

## Physical Resources

The physical environment is heterogeneous and distributed globally. AI Agents can be deployed in data centers, on phones, or embedded in mechanical equipment. Some of the locations may have specialized hardware, such as GPUs or high speed storage. The location may be connected by a wide range of comms infrastructure, such as high speed fiber, cell towers, or satellites, each which has a maximum bandwidth and minimum delay. Also, security functions may be inserted into the communication path, such as firewalls, proxies, and encryption. These resources may fulfill the requirements for only some of the Agent's components and for others the resources may be overkill. So part of the adaptive resolver process is to deploy the components on the "best" resources, but this optimization depends on other factors in the environment.

## Usage Patterns

Usage Patterns involve which functions are called, how frequently, and how much data is transferred. Some properties of a usage pattern inherently give poor performance or consume massive amounts of resources. For example, if the agents need to exchange lots of information then wide-area communication will be difficult. On the other hand, sending small text messages to a remote LLM in a data center, which takes seconds to respond, is totally feasible over a wide-area network. Also, in Agent to Agent communication the sessions are longer and have peer-to-peer interactions, which are not common in the current web-based internet. The comms deployment depends on the usage pattern between agents and has a direct impact on the resulting performance and cost.

## QoS and Security Requirements

The allocation of agents over physical resources has consequences on the performance seen over the communication channel. Some performance properties may be important in a specific context while others not so much. For example, delay is important for real-time control agents, while streaming throughput is needed for training new LLMs. Security and trust may also be traded off against performance. For example, if the interaction is between anonymous read-only agents versus if agents are trusted to perform write operations.

## Cost and resource consumption

The allocation of agents to physical resources also consumes resources. Sometimes the resources are in common and shared. But often the resources are paid for by the owners of the agents. Again, there are tradeoffs between the amount of resources consumed and the performance of the communication channel. All of these properties of the context need to be gathered and their importance agreed upon before the optimization of the communication channel can be started.

# Examples of Adaptive Deployment Modes

The deployment of Agent and comms functionality has lots of possibilities. Here we give examples of broad classes of different deployments. The key features are the location of the agent state, the intensity of communications between components, and the use of specialized hardware. Adaptive deployment needs to support these classes of communication with expectation that new schemes will evolve in the future.

## Client-Server and Dumb Comms

The simplistic model for client-server interaction is that the server is hosted in a data center and the client browser accesses it over the Internet, which is geographically close and has unlimited bandwidth. While this case is common, it does not work for large scale servers with lots of clients. For example millions of fans watching an Indian Cricket match, switching to customized ads all at the same time.

## Server end-point moves closer to client

Large scale web applications have a DNS Adaptive Resolver [[Akamai Global Traffic Management](#)] that attach clients to comms resources that front-end the applications. Effectively, the server's implementation is spread out over the Internet with many distributed end-points and pre-processing being done before the client request reaches the server origin. These same comms services are available to AI Agents, which use standard URLs for their inter-agent comms end points. AI Agents can have additional security and performance services deployed in the comms environment to support scaling their end-points. For example, [Firewall for AI](#) could protect generative AI from prompt attacks, by pre-screening prompts before passing them onto the target agent.

## AI Gateway

Inter-agent communication is not limited to web-based APIs. The communication could be a secure and managed service like a Message bus [RabbitMQ, Kaffka] or an IOT messaging system [ANQP, MQTT]. The AI Gateway could be loaded with functions to help regulate the communication and synchronization with agents that are not on-line at the same time. AI Gateways are especially needed in the case where agents do not trust each other enough to allow incoming requests, such as agents behind a corporate firewall. In that case each agent makes out-going requests to the Gateway's end-points.

## Agent mobility

Agent mobility sends the requesting agent state and code to be closer to the target.
For example a requester agent may rent some computational resources at the same data center as the target, and then physically move its state and code to the target data center, have numerous interactions with the target, and then move its state back to its home resources. While Agent mobility is opposite of moving the server closer to the client, remember that agent comms is peer-to-peer and symmetrical, so this mode is expected.

## Multi-party communication

Agent to Agent comm is not necessarily between only a pair of agents. Many agents may want to communicate over a group chat or conference call. A feature of group communication is that the service can be built around supporting different roles. For example an orchestra pit may have a conductor role along with a tuba and triangle role. Supporting multi-party communication may be necessary to support negotiations where all agents trust the moderator but don't trust each other.

# Adaptive Resolver Architecture

An Adaptive Resolver adds more opportunities for coordination between agents beyond those supported by the Dynamic Resolver. This additional coordination is necessary to allow agents to establish trust and to agree on expectations. As we have seen, the process of deploying a scalable communication channel between agents needs information about the environment's context and the desired requirements and restrictions of the agents themselves. Adaptation is a multi-step process to gather requirements information, optimize a comms specification, and to deploy that specification over the physical resources. Unlike the Dynamic Resolver which is based on Internet DNS, there is no experience running an Adaptive Resolver at scale. Hence, we expect these ideas to evolve rapidly and only give a brief description of the expected functionality.

## Recursive Resolving to get the Authoritative Name Server

The Dynamic Resolver is used to convert the Agent Name into a referral to the Authoritative Name Server that is in charge of the Agent. If either Requester or Target agent requires more to establish trust or feels that the existing comms spec is inadequate, then the Target Agent's Authoritative Name Server returns a Negotiation Invitation to the Requester's Recursive Resolver.
The Authoritative Name Server can act as a broker for the Target and the Recursive Resolver can act as a broker for the Requester. These brokers will carry out the additional step to optimize and deploy the communication channel. (see steps 1,2,3 of figure)

## Requirements Negotiation

The metadata needed for adaptation needs to be gathered and agreed upon. The Dynamic Resolver operation using Target's comms requirement and Requester's context gives a first pass at this metadata. If this meta data is incomplete, then negotiation is done until all parties are satisfied. A yet to be determined algorithm will be used to perform the negotiation and to gather the status of the environment. The metadata should include all the information needed to make the adaptive deployment, including physical resources, expected usage patterns, QoS requirements, and cost restrictions. The output of the negotiations is a Comms Spec (see steps 4,5 of figure)

## Deployment Optimization

Requirements Negotiation defines a Comms Spec that includes the variables, constraints and objective functions for how the communication channel should be set up between agents.  A yet to be determined optimization algorithm will take the Comms Spec as input and return an adequate placement for the agents and comms deployment. The optimization algorithm will most likely start off centralized, but may evolve to be distributed. Entities may have hidden costs and goals, so they may want to perform a sub-optimization on resources they own and only present their sub-results to the group. The output of the deployment is a Placement Spec (see step s6,7 of figure)

## Deployment Setup

The deployment of Agent components and comms must be done by the owner of the resources. The resource owners are given a Placement Spec and report back when their section is complete. When the whole communication channel is complete the Agents are given their end-point URLs and communication can commence (see step 8 of figure)

## Inter-Agent Communication

Well that was a lot of work to get an end-point URL from an Agent Name, but the URL is tailored to the context. All the entities had a chance to include their meta-data and communication was adapted to a wide range of communication modes. The Agents can now interact on a secure, scalable, and inexpensive channel. An additional step is to tear down the channel when the communication is completed. By default this can be done with an inactivity timeout. (see step 9 of figure)

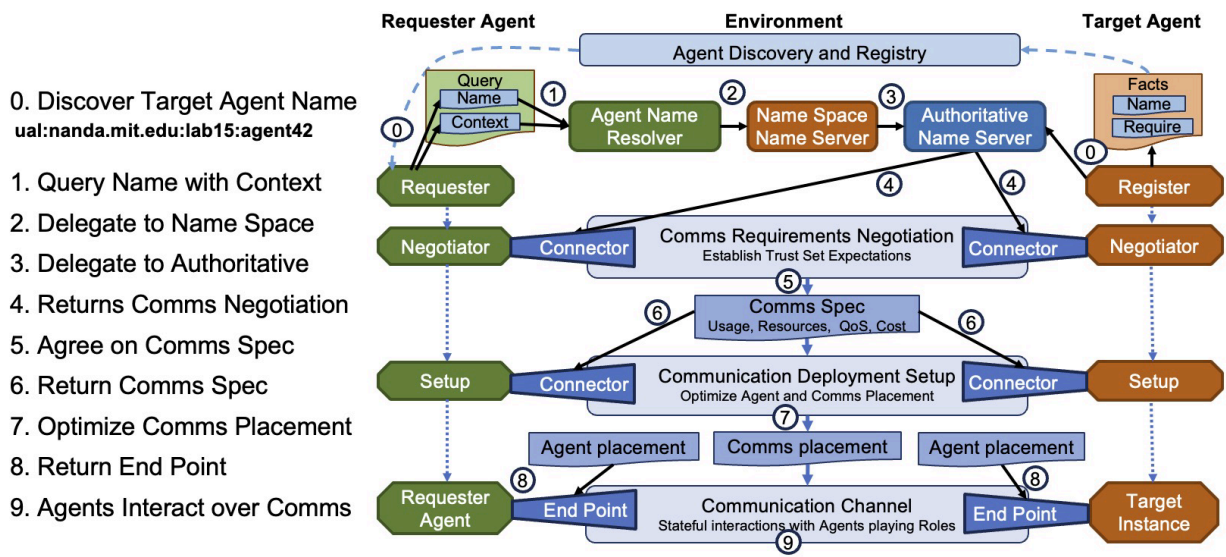

0. Discover Target Agent Name
ual:nanda.mit.edu:lab15:agent42

1. Query Name with Context
2. Delegate to Name Space
3. Delegate to Authoritative
4. Returns Comms Negotiation
5. Agree on Comms Spec
6. Return Comms Spec
7. Optimize Comms Placement
8. Return End Point
9. Agents Interact over Comms

## Future Directions

This document describes the Architecture for the Dynamic Resolution of AI Agent Names. The next step is to refine the Architecture by designing the OpenAPI Specification for the calls between components. The metadata in these calls have only been discussed in the broadest terms. The metadata needs to start off simple and be extensible as we learn more about the actual requirements for the service.

A reference implementation of the service will be essential to test the validity of the architecture. Deployment of this service for a name space can start out centralized, but should quickly include Intermediate Name Servers and Authoritative Name Servers operated by different organizations.

The development should follow industry standards of DevOps for a continuous integration process. Special considerations should be given to stress testing that generates resolution traffic and checks its validity. The Architecture was based on the existing Internet DNS, which has proven to be horizontally scalable and extensible over decades.

## Conclusion

A Dynamic Agent Resolution Service is necessary for a scalable, secure, and efficient Internet of AI Agents. We developed a basic architecture for an AI Agent Name Dynamic Resolver based on the existing Internet DNS recursive resolver. But AI Agent to Agent communication has more opportunities and requirements than the traditional web-based Client-Server communication. AI Agent features will necessitate an evolution to a more comprehensive Adaptive Resolver for Agent Names, where Agents and the comms environment work together to create flexible deployments. But all these architectural requirements are purely speculative. We need to quickly deploy reference implementations to gain experience in the fast moving world of AI Agentic systems.

## Acknowledgements

We would like to thank Ken Huang for his thoughtful suggestions and are looking forward to aligning the Agent Name Service (ANS) and Agent Capability Negotiation and Binding Protocol (ACNBP) with this architecture.

*V0.1, Work in Progress, Request for Comments. Draft*